\documentclass[letterpaper]{article}

\usepackage[T1]{fontenc}
\usepackage[margin=1in]{geometry}
\usepackage{setspace}
\usepackage{amsmath}
\usepackage{amssymb}
\usepackage{graphicx}
\usepackage{float}
\usepackage{placeins}
\usepackage[font=small,labelfont=bf]{caption}
\usepackage[version=4]{mhchem}
\usepackage[super,sort&compress,comma]{natbib}
\usepackage[colorlinks=true,linkcolor=blue,citecolor=blue,urlcolor=blue]{hyperref}
\usepackage{authblk}

\graphicspath{{./}}


\title{Raman-Assisted Multiband Nonlinear Frequency-Conversion Network in a High-\texorpdfstring{$Q$}{Q} LTOI Microdisk}
\author[1]{Zhifan Fang\textsuperscript{\dag}}
\author[1]{Yuxuan He\textsuperscript{\dag}}
\author[1]{Zhangning Pan}
\author[1,2]{Xianfeng Chen}
\author[1]{Yuping Chen\textsuperscript{*}}
\affil[1]{State Key Laboratory of Photonics and Communications, School of Physics and Astronomy, Shanghai Jiao Tong University, Shanghai 200240, China}
\affil[2]{Collaborative Innovation Center of Light Manipulations and Applications, Shandong Normal University, Jinan 250358, China}
\date{\textsuperscript{\dag}These authors contributed equally to this work.\\\textsuperscript{*}Email: \href{mailto:ypchen@sjtu.edu.cn}{ypchen@sjtu.edu.cn}}
\begin{document}

\maketitle

\begin{abstract}
On-chip nonlinear frequency conversion offers a key route to broadband coherent light sources, but spanning telecom, visible, and ultraviolet wavelengths within a single resonator remains challenging. Lithium tantalate-on-insulator (LTOI), which has recently emerged as a promising material platform for integrated photonics, combines strong Raman activity, a large second-order nonlinearity, broad optical transparency and high resistance to photorefractive damage, thereby attracting increasing attention for on-chip nonlinear frequency conversion. Here, we experimentally demonstrate a Raman-assisted multiband frequency-conversion network in a high-\(Q\) LTOI microdisk with a loaded quality factor of \(2.48\times10^6\). The resonant pumping produced high-purity single-mode Raman lasing with a 3.14 mW threshold, 32.44\% slope efficiency, and an excellent side-mode suppression ratio (SMSR) of 29.5 dB. Under a nearby pump condition, we also observe multiple Stokes components together with an anti-Stokes line on the short-wavelength side of the pump. The resulting multiple intracavity Stokes fields subsequently acted as frequency seeds for cascaded \(\chi^{(2)}\) processes, producing near-infrared and visible signals and extending the emission to 312.6 nm in the ultraviolet. These findings establish the cooperative action of Raman gain and second-order nonlinearity across widely separated spectral bands within a single microcavity. The device therefore offers a route toward integrated multiband light sources and a platform for studying coupled nonlinear dynamics.
\end{abstract}

\section{Keywords}
integrated nonlinear photonics, lithium tantalate-on-insulator, stimulated Raman scattering,\\
cascaded frequency conversion, ultraviolet generation

\section{Introduction}\label{sec:introduction}
Integrated nonlinear photonics provides a scalable route to chip-scale optical generation, manipulation, and signal processing, with applications in coherent light sources, high-capacity communications, quantum technologies, and photonic computing \cite{ref1,ref2,ref3,ref4,ref5}. To fully unlock the potential of these integrated platforms, high-$Q$ optical microcavities have emerged as key components for nonlinear frequency conversion, as their small mode volumes and long photon lifetimes enhance intracavity fields, increase effective interaction lengths, and reduce nonlinear thresholds. Their dense mode spectra can also support multiple resonant fields, enabling cascaded processes that connect widely separated optical frequencies \cite{ref6,ref7,ref8,ref9,ref10}. Nevertheless, the nonlinear performance and accessible spectral range depend not only on their geometry, dispersion, and mode structure, but also critically on the material hosting the optical modes. An effective nonlinear photonic platform must combine strong optical nonlinearity, low optical loss, and compatibility with reliable nanofabrication. In this context, lithium tantalate-on-insulator (LTOI) has recently emerged as a promising wafer-scale platform due to its strong second-order and electro-optic responses, relatively low birefringence, a broad transparency window, and substantially improved resistance to photorefractive damage \cite{ref11,ref12,ref13,ref14}. These properties have enabled rapid progress in high-speed electro-optic modulators, Kerr microcombs, periodically poled frequency converters, and visible-to-ultraviolet second-harmonic generation \cite{ref15,ref16,ref17,ref18,ref19}. Despite these advances, the strong Raman response intrinsic to lithium tantalate has not yet been comparably exploited in integrated LTOI devices. In fact, Raman processes have been extensively harnessed for low-threshold Raman lasing, wavelength-tunable and multiband Raman emission, Raman--Kerr interactions, and Raman-assisted optical upconversion in lithium niobate-on-insulator (LNOI), a closely related ferroelectric platform \cite{ref20,ref21,ref22,ref23}. These results show that Raman scattering can serve not merely as a parasitic process, but as an active gain mechanism for generating and connecting new optical frequencies. However, the stronger photorefractive response and increasing absorption of lithium niobate toward the near-ultraviolet can constrain high-intracavity-power operation and multistep conversion toward shorter wavelengths \cite{ref14,ref24,ref25}. By comparison, lithium tantalate offers a higher photorefractive-damage threshold and a transparency window extending from approximately 0.28 \(\mu\)m in the ultraviolet to 5.5 \(\mu\)m in the mid-infrared \cite{ref11,ref26,ref27}, making LTOI particularly attractive for combining Raman gain with cascaded \(\chi^{(2)}\) interactions. To the best of our knowledge, cavity-enhanced Raman gain has not previously been systematically harnessed in an integrated LTOI resonator to realize Raman lasing and Raman-assisted multiband nonlinear frequency conversion.\looseness=-2

\begin{figure}[!t]
\centering
\includegraphics[width=0.88\textwidth]{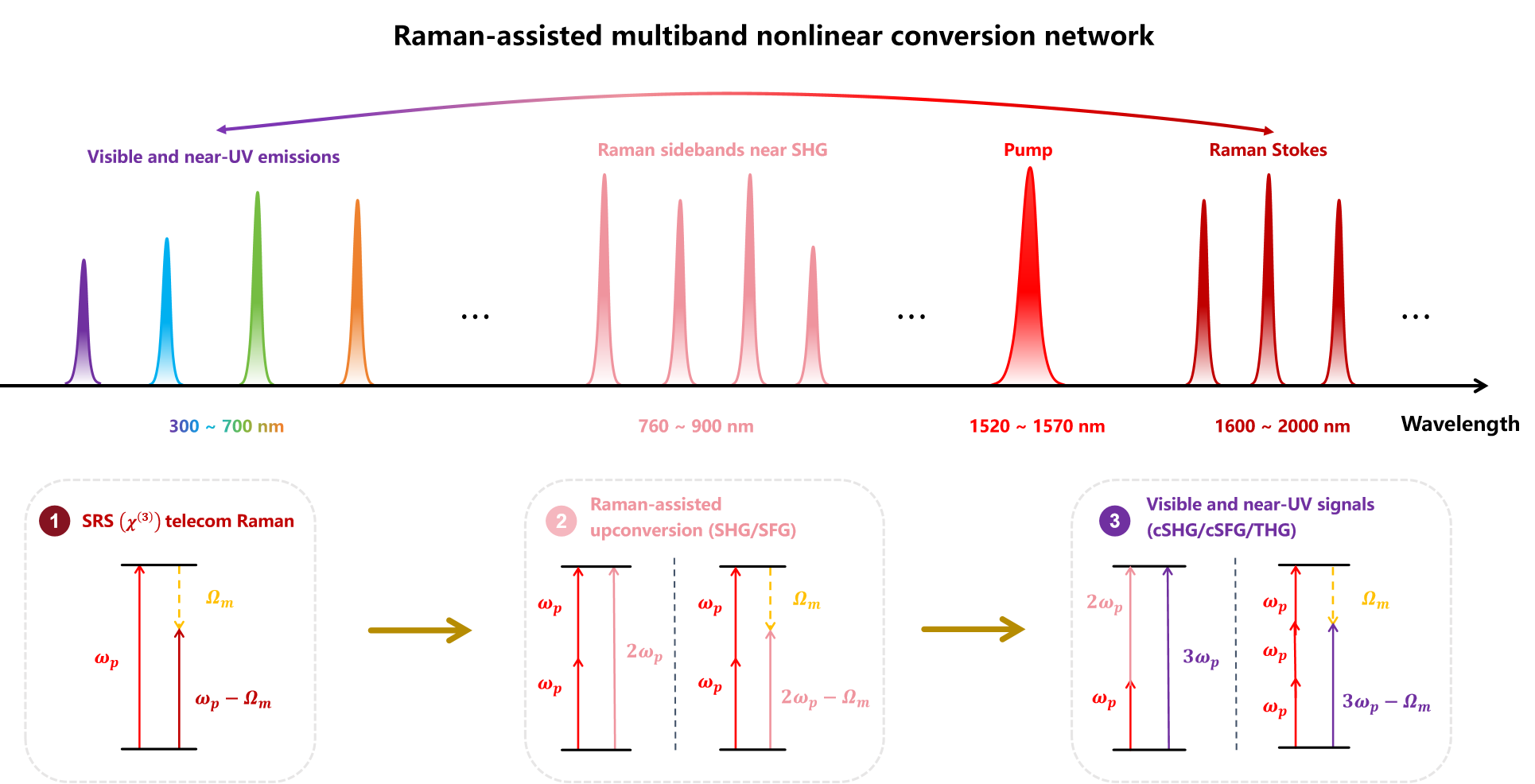}
\caption{Schematic of the Raman-assisted multiband nonlinear frequency-conversion network in a LTOI microdisk. Telecom-band resonant pumping generates Raman Stokes fields, which subsequently seed cascaded $\chi^{(2)}$ upconversion and harmonic generation, producing signals from the near-infrared to the ultraviolet.\label{fig:network}}
\end{figure}

Here, we address this gap by experimentally characterizing a multiband frequency-conversion network in a LTOI microdisk with \(Q \approx 2.48 \times 10^{6}\). By tuning the pump to high-\(Q\) cavity modes, we observe multiple Stokes components together with an anti-Stokes line on the short-wavelength side of the pump. Selective excitation of a Raman mode yields high-purity, single-mode Raman lasing in the telecom band with a low threshold of 3.14 mW, a high slope efficiency of 32.44\%, and an SMSR of 29.5 dB. Subsequently, the resulting multiple Stokes fields served as intracavity seeds for Raman-assisted \(\chi^{(2)}\) upconversion. In the near-infrared upconversion region, we observed efficient second-harmonic generation alongside Raman-assisted \(\chi^{(2)}\) mixing, and the 811.3 nm line reached a normalized conversion efficiency of 0.571\% mW$^{-1}$ with a 22.5 dB signal-to-background ratio. Further cascaded conversion generated orange-red, green, blue, and 312.6 nm ultraviolet emission from the same microdisk. Together, these results elucidate the complex interplay between multiple Raman-active phonon branches and cascaded nonlinear processes in LTOI. Furthermore, they experimentally validate the critical advantage of LTOI's broad transparency window for facilitating nonlinear evolution across widely separated spectral bands, especially toward the ultraviolet. This work establishes a significant experimental platform for developing broadband integrated coherent sources spanning from the ultraviolet to the infrared, and opens new avenues for exploring multiband on-chip nonlinear dynamics.

\section{Results and discussion}\label{sec:results}

\subsection{Fabrication and Characterization of the High-\texorpdfstring{$Q$}{Q} LTOI Microdisk}\label{sec:fabrication}

\begin{figure}[!t]
\centering
\includegraphics[width=0.78\textwidth]{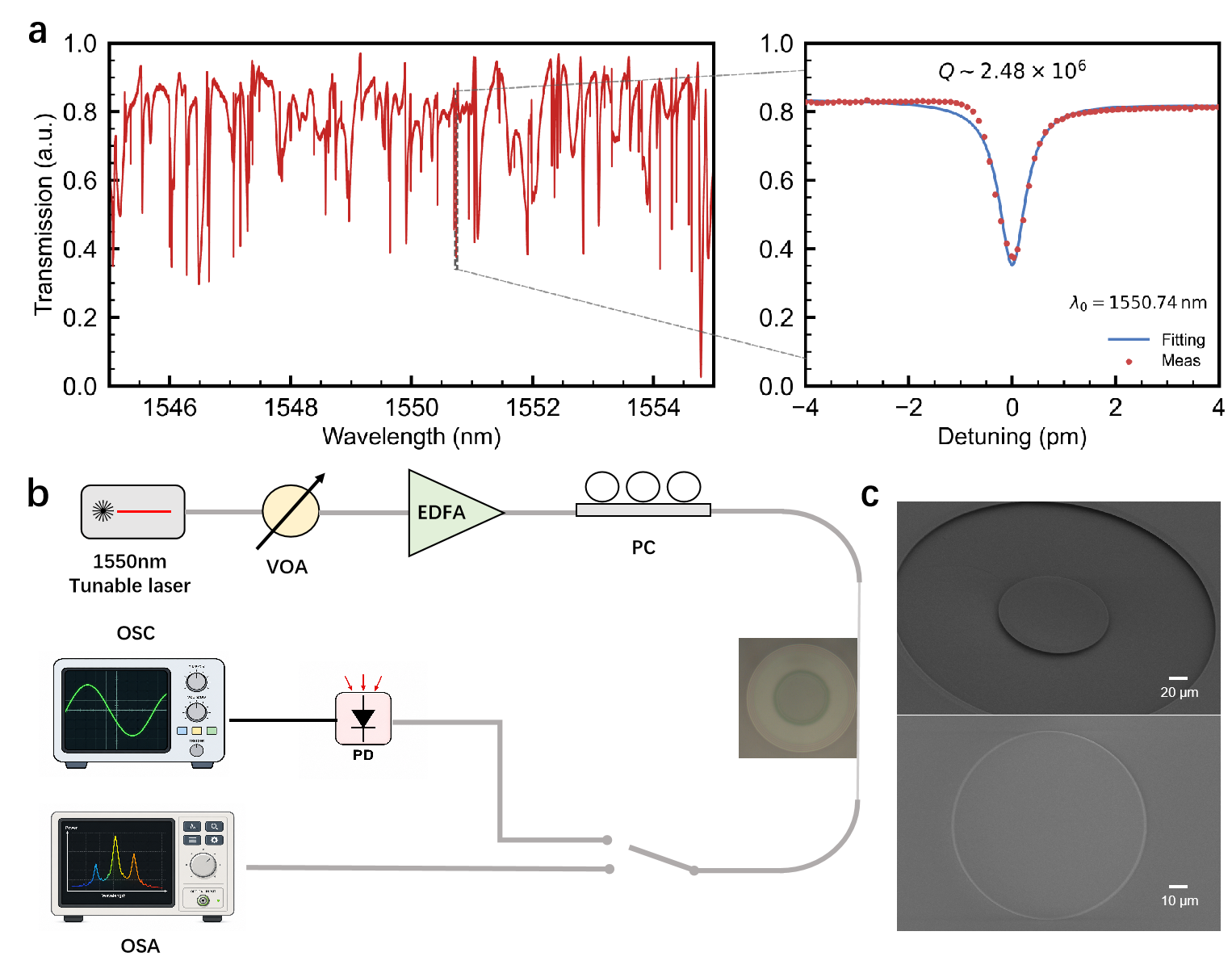}
\caption{Device characterization and experimental setup. \textbf{a}, Normalized transmission spectrum near 1550 nm. Inset, a Lorentzian fit of a whispering-gallery mode at 1550.74 nm indicating a loaded $Q$-factor of $2.48\times10^{6}$. \textbf{b}, Schematic of the optical measurement setup. VOA, variable optical attenuator; EDFA, erbium-doped fiber amplifier; PC, polarization controller; PD, photodetector; OSC, oscilloscope; OSA, optical spectrum analyzer. \textbf{c}, Scanning electron microscope images of the LTOI microdisk.\label{fig:device}}
\end{figure}

The x-cut LTOI wafer comprised a 600-nm lithium tantalate film, a 4.7-$\mu$m SiO$_2$ buffer layer, and a 675-$\mu$m silicon substrate. We fabricated a microdisk with a radius of 104.8~\(\mu\mathrm{m}\) by photolithography-assisted chemo-mechanical polishing \cite{ref28,ref29}. A detailed fabrication procedure and the corresponding process flow are provided in Supplementary Note 1 and Fig. S1. We first characterized the high-Q response of the microdisk in the pump band. The input power was kept low enough to suppress intracavity thermal distortion and preserve a Lorentzian resonance lineshape. Scanning the pump from 1540 to 1565 nm revealed numerous optical modes, with a representative interval shown in Figure~\ref{fig:device}a. This resonance density provided multiple candidate modes for pump tuning in the subsequent nonlinear experiments. The inset resolves a representative whispering-gallery mode near 1550.7 nm. A Lorentzian fit yielded a loaded Q of \(2.48 \times 10^{6}\). Figure~\ref{fig:device}c shows the scanning electron microscope images of the LTOI microdisk. The resulting microdisk possesses smooth top and sidewall surfaces, which are essential for suppressing scattering loss, increasing the intracavity circulating power, and therefore reducing the stimulated-Raman-scattering threshold.

Figure~\ref{fig:device}b shows the experimental setup. A narrow-linewidth tunable continuous-wave laser near 1550 nm provided the pump. A variable optical attenuator (VOA) controlled the power entering the erbium-doped fiber amplifier (EDFA). A polarization controller (PC) then aligned the pump polarization with a high-Q microdisk mode. The pump was coupled through a tapered optical fiber fabricated by the heat-and-pull method \cite{ref30}. Its 1--2 $\mu$m waist improved spatial overlap between the fiber mode and the whispering-gallery modes. A three-axis translation stage controlled the taper-microdisk gap, while an optical microscope enabled real-time monitoring of the coupling region. Light collected by the same tapered fiber was directed to either of two detection paths. A photodetector (PD) and oscilloscope recorded transmission during wavelength scans to locate resonances and evaluate coupling. An optical spectrum analyzer measured the pump band and selected features near the second harmonic. A fiber-coupled spectrometer recorded most near-infrared Raman-assisted upconversion features and visible/ultraviolet spectra from 200 to 1100 nm. It should be emphasized that, owing to the different spectral response ranges, spectral resolutions, and detection sensitivities of the spectrometers used in this work, the comparison of signals across different wavelength regions is mainly based on their spectral positions, frequency-shift correlations, and power-dependent behaviors. Direct quantitative comparison of absolute intensities measured by different instruments is therefore avoided.

\subsection{Efficient Stimulated Raman Scattering and Single-Mode Raman Lasing}\label{sec:raman}

\begin{figure}[!t]
\centering
\includegraphics[width=0.74\textwidth]{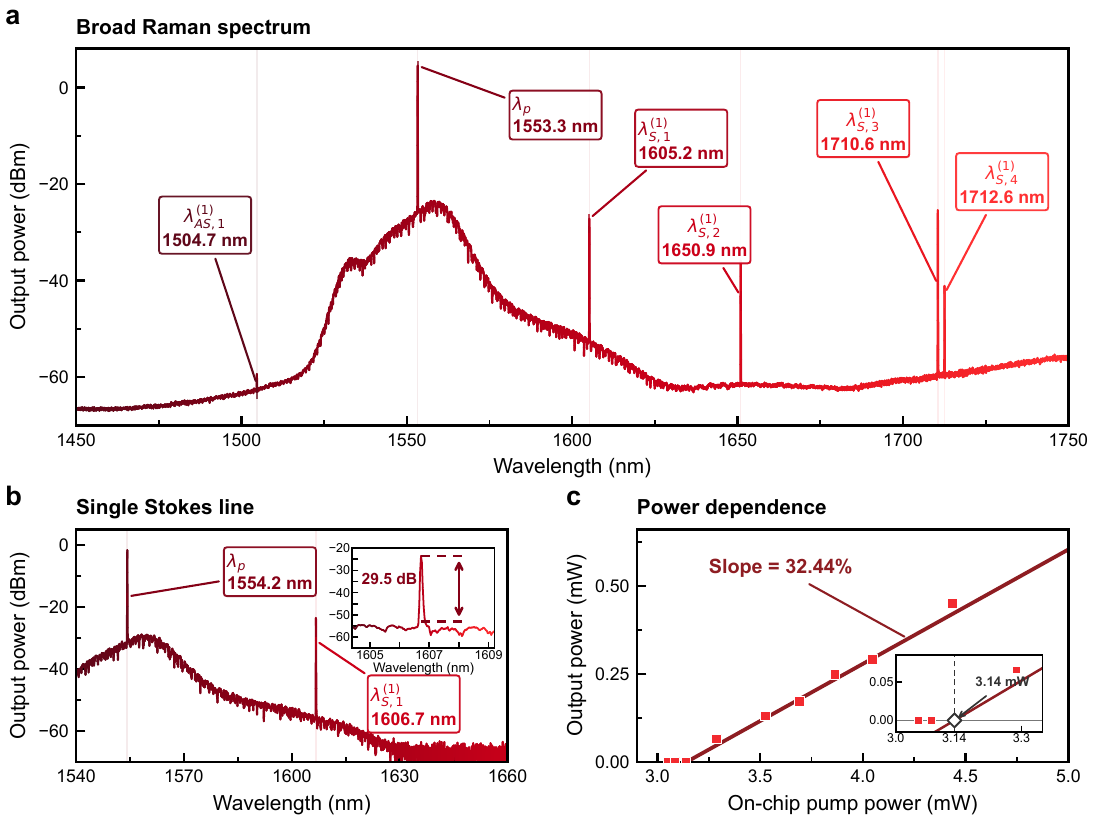}
\caption{Stimulated Raman scattering and single-mode lasing. \textbf{a}, Broadband Raman spectrum generated by a 1553.3 nm pump, showing primary Stokes and anti-Stokes lines. \textbf{b}, Single-mode Raman lasing at 1606.7 nm (1554.2 nm pump) with a 29.5 dB side-mode suppression ratio. \textbf{c}, Power dependence of the single-mode Raman laser, showing a 3.14 mW threshold and 32.44\% slope efficiency.\label{fig:raman}}
\end{figure}

We next examined stimulated Raman scattering in the telecom band. With the pump coupled to a high-\(Q\) resonance at 1553.3 nm above threshold, the spectrum contained four prominent Stokes lines [Figure~\ref{fig:raman}a]. These lines appeared at \(\lambda_{S,1}^{(1)} =\)1605.2 nm, \(\lambda_{S,2}^{(1)} =\)1650.9 nm, \(\lambda_{S,3}^{(1)} =\)1710.6 nm, and \(\lambda_{S,4}^{(1)} =\)1712.6 nm, where the superscript denotes the Stokes order. Using \(\Delta\tilde{\nu}=10^7(1/\lambda_p-1/\lambda_S)\), where \(\lambda_{p}\) and \(\lambda_{S}\) are the pump and Stokes wavelengths, respectively, we obtained Raman shifts of 208, 381, 592, and 599 \(\mathrm{cm}^{-1}\). The first two shifts were assigned to the \(A_{1}(\mathrm{TO})_{1}/E(\mathrm{TO})_{2}\) and \(E(\mathrm{TO})_5\) branches \cite{ref31,ref32}. The two higher shifts were consistent with the closely spaced \(A_1(\mathrm{TO})_4/E(\mathrm{TO})_8\) branches \cite{ref31,ref32}. Notably, an anti-Stokes line at \(\lambda_{AS,1}^{(1)} =\)1504.7 nm also appeared on the short-wavelength side of the pump. Its shift, \(\Delta\nu \approx\)208 \(\mathrm{cm}^{-1}\), matched that of \(\lambda_{S,1}^{(1)}\) and the \(A_1(\mathrm{TO})_1/E(\mathrm{TO})_2\) branch. Conventional anti-Stokes scattering requires a pump photon to absorb lattice vibrational energy and is therefore limited by the thermal phonon population \cite{ref33,ref34}. Above threshold, however, the intense intracavity Stokes field may also enable cavity-enhanced Raman-assisted four-wave mixing \cite{ref35,ref36}. This process satisfies \(\omega_{AS} = 2\omega_{p} - \omega_{S} = \omega_{p} + \Omega_{R}\), where \(\Omega_{R}\) is the relevant phonon frequency.

To quantify the Raman-lasing performance, we finely adjusted the pump wavelength to select a pump--Stokes resonance pair that favored a single Raman lasing line, enabling the measurement of its lasing threshold and slope efficiency. At a pump wavelength of 1554.2 nm, the 1606.7 nm Raman line dominated the spectrum [Figure~\ref{fig:raman}b]. Competing Raman lines were suppressed, yielding an SMSR of 29.5 dB. Figure~\ref{fig:raman}c plots the Raman output against on-chip pump power. Here, we define on-chip pump power as the power coupled into the microdisk mode after correcting for fiber transmission and taper-coupling losses. Below threshold, the Raman signal remained at the noise floor; above threshold, it increased linearly with on-chip pump power. A linear fit of the experimental data revealed a low threshold of 3.14 mW and a high slope efficiency of 32.44\%. Such highly efficient lasing dynamics are fundamentally underpinned by the exceptional \(Q\)-factor of the LTOI microdisk, which provides strong resonant enhancement for both the pump and the localized Raman fields.

It should be noted that Figure~\ref{fig:raman}a includes only the Raman lines captured within the available detection window and coupling condition. Because the microdisk supports several Raman-active phonon modes, the pump can generate first-order Stokes fields through both low- and high-frequency phonons. These primary Stokes fields can then act as secondary Raman pumps and generate higher-order components through cascaded scattering \cite{ref37,ref38}. Although some components generated via high-frequency phonons or deep cascaded scattering shift to longer wavelengths beyond the spectrometer's detection range, the intracavity ensemble of these fields can still serve as critical internal frequency seeds for the multiband nonlinear conversion network, as will be demonstrated in the following sections.\looseness=-1

\subsection{Raman-Assisted Quadratic Lasing and Nonlinear Upconversion}\label{sec:upconversion}

\begin{figure}[!t]
\centering
\includegraphics[width=0.70\textwidth]{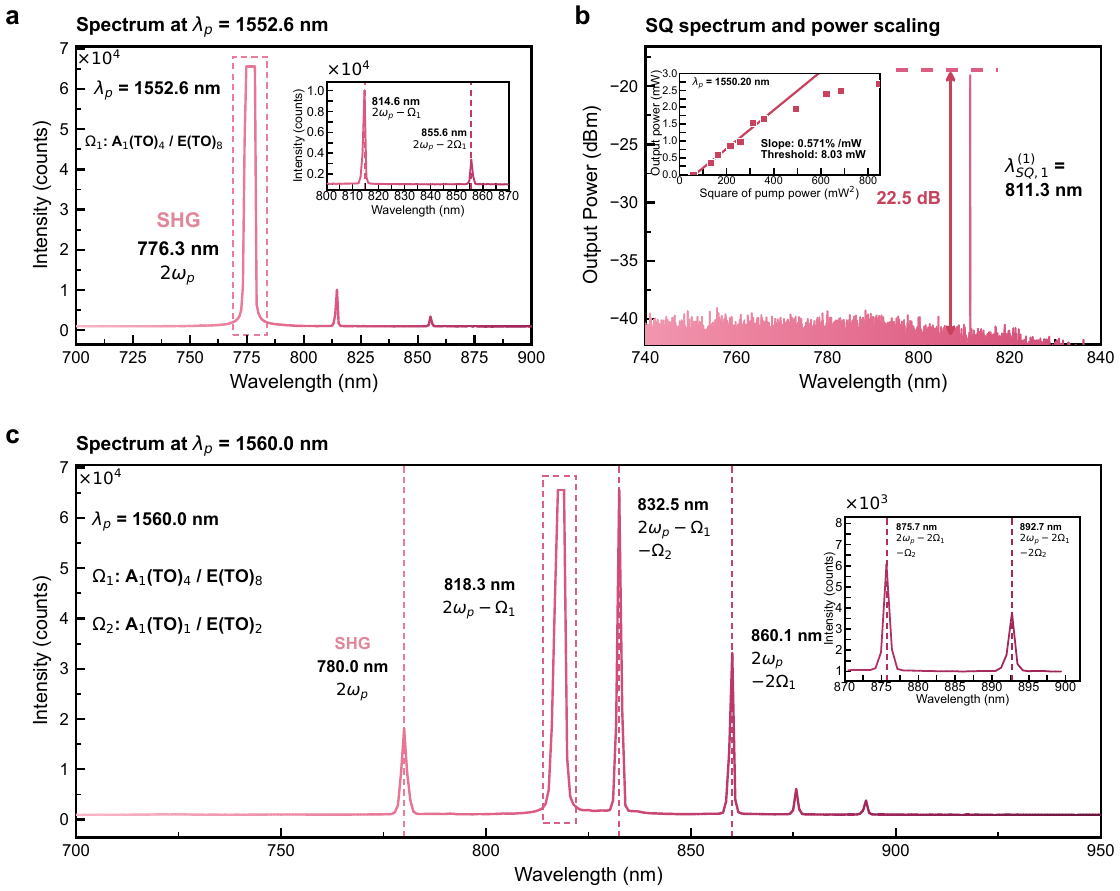}
\caption{Near-infrared Raman-assisted nonlinear upconversion. \textbf{a}, Spectrum near the second-harmonic band with a 1552.6 nm pump. Inset, sum-frequency and second-harmonic generation signals of primary Raman Stokes fields. \textbf{b}, Isolated Raman-assisted sum-frequency line at 811.3 nm under 1550.2 nm pumping. The peak shows a 22.5 dB signal-to-background ratio; the inset plots the output power against the square of the pump power, giving a threshold of 8.03 mW and a normalized conversion efficiency of 0.571\% mW$^{-1}$. \textbf{c}, Upconversion spectrum (1560.0 nm pump) resulting from cascaded processes across two distinct Raman-active phonon branches.\label{fig:nir}}
\end{figure}

The telecom-band Stokes fields and single-mode Raman laser established substantial intracavity Raman gain. We next combined this gain with the intrinsic second-order nonlinearity of LTOI to examine near-infrared Raman-assisted lasing and upconversion. To precisely characterize the initially weak upconverted signals at a relatively low pump power ($\sim$ 4 mW), we employed a high-sensitivity fiber-coupled spectrometer (200--1100 nm) combined with a 780/1550-nm wavelength-division multiplexer to actively suppress the pump-induced background. Pumping a 1552.6 nm resonance generated a pronounced SHG signal at 776.3 nm (\(2\omega_p\)) [Figure~\ref{fig:nir}a]. The flat-topped peak resulted from spectrometer saturation rather than the intrinsic spectral profile. Two weaker lines at 814.6 and 855.6 nm appeared on the long-wavelength side of the SHG peak. Their wavenumber offsets were approximately 606 and 1194 \(\mathrm{cm}^{-1}\), respectively. We denote the phonon frequency of the \(A_1(\mathrm{TO})_4/E(\mathrm{TO})_8\) branch as \(\Omega_{1}\). The two positions were consistent with sum-frequency generation (SFG) between the pump and first-order Raman Stokes field (\(2\omega_p-\Omega_{1}\)) and SHG of that Stokes field (\(2\omega_p-2\Omega_{1}\)), respectively.

We then optimized the pump wavelength and power to isolate a dominant SFG line at 811.3 nm. Figure~\ref{fig:nir}b shows the line with a 22.5 dB signal-to-background ratio and its power dependence. Unlike the linear power scaling characteristic of the first-order Raman scattering, this upconversion process reflects a Raman-seeded quadratic process, featuring an on-chip pump threshold of 8.03 mW. Above this threshold, the output power scaled quadratically with the on-chip pump power, achieving a high normalized conversion efficiency of 0.571\% \(\mathrm{mW}^{-1}\). Above approximately 20 mW, the output deviated from quadratic scaling and began to saturate, a behavior that likely arises from thermally induced resonance shifts, pump depletion, and gain competition among multiple Raman modes \cite{ref39,ref40,ref41}.

Motivated by the inherent multimode characteristics of the LTOI microdisk, we subsequently tuned the pump wavelength at 1560.0 nm to access a broader set of spectral lines. Under this condition, direct SHG at 780.0 nm was no longer dominant, and discrete lines appeared from 775 to 900 nm [Figure~\ref{fig:nir}c]. The 818.3 nm (\(2\omega_p-\Omega_1\)) and 860.1 nm (\(2\omega_p-2\Omega_1\)) lines were shifted from SHG by approximately 600 and 1194 cm$^{-1}$, respectively. Their agreement with the shifts observed under 1552.6 nm pumping showed that the \(A_1(\mathrm{TO})_4/E(\mathrm{TO})_8\) branches continued to participate in Raman-assisted upconversion. Furthermore, three additional lines revealed more complex cascaded processes involving a second Raman-active phonon branch, denoted as \(\Omega_2\) (\(A_1(\mathrm{TO})_1/E(\mathrm{TO})_2\)). The 832.5 nm line (\(2\omega_p-\Omega_1-\Omega_2\)) was displaced by approximately 208 \(\mathrm{cm}^{-1}\) from the 818.3 nm peak. This position was consistent with SFG involving the pump and a cascaded second-order Stokes field. The inset further resolves lines at 875.7 nm (\(2\omega_p-2\Omega_1-\Omega_2\)) and 892.7 nm (\(2\omega_p-2\Omega_1-2\Omega_2\)). Their shifts from the 860.1 nm peak were approximately 207 and 424 \(\mathrm{cm}^{-1}\), respectively. These lines were consistent with SFG between first- and second-order Stokes fields and SHG of the second-order Stokes field, respectively.

Specifically, once generated and resonantly enhanced, these upconverted lines can also act as additional seeds for cascaded sum-frequency generation, harmonic generation, and other Raman-assisted nonlinear processes. The conversion near the second-harmonic band therefore forms an intermediate stage linking telecom-band stimulated Raman scattering to visible and ultraviolet emission.

\subsection[Raman-Assisted Cascaded Nonlinear Conversion and Visible/Ultraviolet Emission]{Raman-Assisted Cascaded Nonlinear Conversion and Visible/Ultraviolet\\ Emission}\label{sec:visible-uv}

Building on the near-infrared upconversion, we examined visible and ultraviolet emission at higher pump powers and under different resonance conditions. By adjusting the pump wavelength, pump power, polarization state, and taper--microdisk coupling, we observed four principal emission bands in the orange-red, green, blue, and ultraviolet spectral regions. Because the present setup did not include a calibrated visible-band power-collection and detection chain, the collection efficiency, transmission loss, and detector responsivity varied substantially among different wavelength bands. Therefore, in this section, we focus on identifying the dominant frequency-conversion channels and their pump-power scaling behavior, rather than quantitatively comparing the absolute conversion efficiencies across different spectral regions.

\begin{figure}[!t]
\centering
\includegraphics[width=0.72\textwidth]{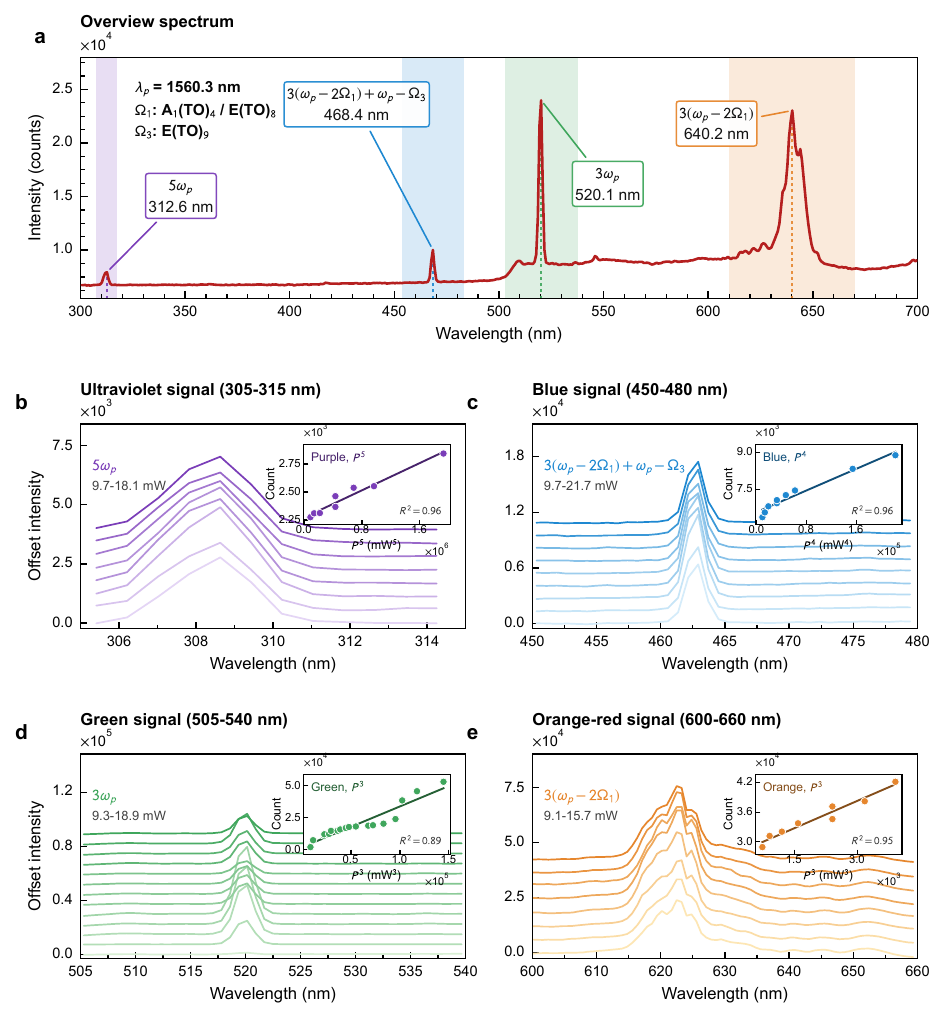}
\caption{Visible and ultraviolet emission from cascaded nonlinear conversion. \textbf{a}, Broadband spectrum (1560.3 nm pump) displaying orange-red, green, blue, and near-ultraviolet emission. \textbf{b--e}, Spectral evolution and power dependence (insets) for the ultraviolet (\textbf{b}), blue (\textbf{c}), green (\textbf{d}), and orange-red (\textbf{e}) signals, showing fifth-, fourth-, third-, and third-order power scaling, respectively.\label{fig:visible}}
\end{figure}

Figure~\ref{fig:visible}a shows a representative broadband spectrum obtained with a pump wavelength of 1560.3 nm. Four main spectral groups are observed at approximately 640.2 nm, 520.1 nm, 468.4 nm, and 312.6 nm, corresponding to orange-red, green, blue, and ultraviolet emission, respectively. The green line at 520.1 nm is consistent with cascaded third-harmonic generation through sequential SHG and SFG by the intrinsic second-order nonlinearity of LTOI. In this process, the pump first generates its second harmonic, and the generated second-harmonic field subsequently mixes with the residual pump field through sum-frequency generation, producing the \(3\omega_p\) component. The ultraviolet signal at 312.6 nm lies near the fifth-harmonic band of the pump which is most plausibly assigned to a higher-order cascaded process, such as sum-frequency generation between the second- and third-harmonic fields, giving rise to a \(5\omega_p\) component. 

In contrast, the orange-red and blue signals did not coincide with direct pump harmonics and instead indicated Raman-assisted cascaded conversion. The 640.2 nm signal could originate from the third harmonic of an inferred intracavity field near 1.92~\(\mu\mathrm{m}\). This wavelength is consistent with a second-order Stokes field generated through cascaded Raman scattering, with an overall Raman shift of approximately 1202 cm$^{-1}$, close to twice the frequency shift of the \(A_1(\mathrm{TO})_4/E(\mathrm{TO})_8\) Raman-active phonon branch. Cascaded Raman scattering can first generate the second-order Stokes field, which is then upconverted into the orange-red band. Its broad profile is also consistent with the adjacent telecom-band Stokes lines \(\lambda_{S,3}^{(1)}\) and \(\lambda_{S,4}^{(1)}\) [Figure~\ref{fig:raman}a]. The weaker blue line at 468.4 nm is compatible with SFG involving the orange-red field and an inferred Stokes component near 1745 nm, which is associated with the Raman-active $E(\mathrm{TO})_9$ phonon \cite{ref32}. Although this Stokes component did not manifest as a predominant Raman peak in our earlier telecom-band experiments, it may be resonantly enhanced under the present pump wavelength and coupling condition and serve as a weak internal seed for sum-frequency generation. Consequently, the emergence of the blue emission further suggests that the distinct Raman Stokes components within the microdisk are not isolated entities. Rather, they actively interact via cascaded nonlinear mixing to synthesize higher-frequency outputs.

To further examine these assignments, we measured the pump-power-dependent spectra for each principal emission band. Figures~\ref{fig:visible}(b--e) show the power-dependent ultraviolet (305-315 nm), blue (450-480 nm), green (505-540 nm), and orange-red (600-660 nm) spectra. Specifically, as shown in the insets, the green emission around 520 nm increases monotonically with pump power and shows an approximately cubic scaling, with the fitted dependence \(R^{2}=0.89\). The orange-red band also displays an overall cubic dependence on pump power, with \(R^{2}=0.95\). The blue emission exhibits a higher-order scaling behavior. As the pump power increases from 9.7 to 21.7 mW, the blue peak in the 450--480 nm window becomes progressively stronger, and the fitted intensity follows an approximate fourth-order dependence, with \(R^{2}=0.96\). A similar higher-order trend is also observed for the ultraviolet signal in the 305--315 nm window, which gradually emerges from the background and exhibits an approximate fifth-order dependence on pump power with \(R^{2}=0.96\). These power-law trends support the assignment of visible and ultraviolet emission to cascaded nonlinear conversion pathways, although their absolute intensities remain sensitive to resonance detuning, mode overlap, and the power-dependent buildup of intermediate fields.

\section{Conclusion}\label{sec:conclusion}

In summary, by exploiting cavity enhancement and the strong Raman activity, we have demonstrated for the first time a multiband nonlinear frequency conversion network within a LTOI microdisk that combines Raman lasing, cascaded Raman scattering, harmonic generation, and multistep \(\chi^{(2)}\) processes. The high-Q microdisk not only supports efficient generation of high-purity Raman laser emission but also resonantly enhances the pump and excites multiple Raman-active phonons. The resulting Stokes components are subsequently upconverted through the strong second-order nonlinearity of lithium tantalate into the second-, third-, and higher-harmonic spectral regions. This behavior highlights the complementary roles of Raman and second-order nonlinearities in LTOI and establishes thin-film lithium tantalate as an efficient on-chip frequency-conversion platform that supports cooperative nonlinear processes rather than only isolated SHG or Raman-lasing functions. This work therefore provides a physical basis and experimental framework for engineering Raman-assisted cascaded nonlinear dynamics in emerging LTOI photonic devices. Further improvements in dispersion engineering, mode matching, visible-light power calibration, and phase-matching design should enhance the conversion efficiency, spectral selectivity, and device-to-device reproducibility.

\section{Associated Content}
\subsection{Supporting Information}
The Supporting Information includes the fabrication process, modal-phase-matching considerations, Raman peak assignments, and cascaded frequency-conversion pathway analysis, including Fig. S1 and Tables S1 and S2.

\subsection{Data Availability Statement}
The data that support the findings of this study are available from the corresponding author upon reasonable request.

\section{Author Information}
\subsection{Corresponding Author}
Yuping Chen -- State Key Laboratory of Photonics and Communications, School of Physics and Astronomy, Shanghai Jiao Tong University, Shanghai 200240, China; Email: \href{mailto:ypchen@sjtu.edu.cn}{ypchen@sjtu.edu.cn}

\subsection{Authors}
Zhifan Fang -- State Key Laboratory of Photonics and Communications, School of Physics and Astronomy, Shanghai Jiao Tong University, Shanghai 200240, China\\
Yuxuan He -- State Key Laboratory of Photonics and Communications, School of Physics and Astronomy, Shanghai Jiao Tong University, Shanghai 200240, China\\
Zhangning Pan -- State Key Laboratory of Photonics and Communications, School of Physics and Astronomy, Shanghai Jiao Tong University, Shanghai 200240, China\\
Xianfeng Chen -- State Key Laboratory of Photonics and Communications, School of Physics and Astronomy, Shanghai Jiao Tong University, Shanghai 200240, China; Collaborative Innovation Center of Light Manipulations and Applications, Shandong Normal University, Jinan 250358, China

\section{Author Contributions}
\textsuperscript{\dag}Zhifan Fang and Yuxuan He contributed equally to this work.

\section{Funding}
This work was supported by the National Natural Science Foundation of China (12134009 and 12474335).

\section{Notes}
The authors declare no competing financial interest.

\section{Acknowledgments}
The authors thank Shanghai Novel Si Integration Technology Co., Ltd. (NSIT) for providing the wafers.
The authors thank Prof. Ya Cheng, Prof. Jintian Lin and Renhong Gao for their assistance with chemo-mechanical polishing. The authors thank
the Center for Advanced Electronic Materials and Devices of Shanghai Jiao Tong University for its support in device fabrication.

\FloatBarrier

\end{document}